\documentclass{mn2e}

\usepackage[dvips]{graphicx}

\newcommand{\kms}{\mbox{ km s$^{-1}$}}

\newcommand{\msun}{\mbox{ M$_\odot$}}

\newcommand{\be}{\begin{equation}}
\newcommand{\ee}{\end{equation}}

\def\ltsima{$\; \buildrel < \over \sim \;$}
\def\simlt{\lower.5ex\hbox{\ltsima}}
\def\gtsima{$\; \buildrel > \over \sim \;$}
\def\simgt{\lower.5ex\hbox{\gtsima}}

\begin{document}

\title[Metallicity relations  and Hierarchical clustering]{Fingerprints of the Hierarchical Building up  of  the Structure on  the Mass-Metallicity Relation }

\author[Tissera et al.]{Patricia B. Tissera$^{1,2}$, Mar\'{\i}a E. De
Rossi$^{1,2}$ and Cecilia Scannapieco $^{1,2}$\\
$^1$  Consejo Nacional de Investigaciones Cient\'{\i}ficas
y T\'ecnicas, CONICET, Argentina.\\ 
$^2$ Instituto de Astronom\'{\i}a y F\'{\i}sica del Espacio, Casilla de Correos 67,
Suc. 28, 1428, Buenos Aires, Argentina.\\
}

\maketitle

\begin{abstract}

We study the mass-metallicity relation of galactic systems with stellar masses larger than  $10^9 \msun h^{-1}$
in $\Lambda$CDM scenarios  by using  chemical hydrodynamical simulations.
We find that this  relation arises naturally as a consequence of the formation of the structure in a hierarchical scenario. 
The hierarchical building up of the structure determines a characteristic stellar mass at  $M_c \approx  10^{10.2} \msun h^{-1}$
which exhibits approximately  solar metallicities from $z \approx 3$ to $z=0$.
This characteristic mass separates galactic systems in two groups with massive ones forming most of
their stars and metals  at high redshift. We find evolution in the zero point and slope of  the 
  mass-metallicity relation driven mainly by 
the low mass systems which exhibit the larger variations in the chemical properties.
Although stellar mass and circular velocity are directly related, the correlation  between circular velocity
and metallicity shows a larger evolution with redshift making this relation more appropriate to  confront  models
and observations.
The dispersion found in both relations
is a function of the stellar mass and reflects the different dynamical history of evolution of the systems.
%The simulated mass-metallicity relation and the effective yields suggest the need for strong Supernova 
%feedback principally at the low mass end.

\end{abstract}


\section{Introduction}
\label{intro}

Our knowledge of the chemical evolution of the Universe has improved dramatically in the last years
as the result of  the high precision data gathered on local and high redshift galaxies and
on the interstellar and intergalactic media. 
%(e.g. Lilly, Carollo \& Stockton 2003; Kobulnicky et al. 2003; Pettini et al. 1998, 2001;
 %Kobulnicky \& Koo 2000; Mehlert et al. 2002; Lemoine-Busserolle et al. 2003, Erb et al. 2003).
 In particular, estimates of the evolution of 
the well-known local luminosity-metallicity relation  suggest a change in the zero point and
slope so that at a given metallicity, systems are brighter in the past \cite{kk04,maier04,sha04}. 
Lequeux et al. (1979) first claimed that the fundamental relation was 
actually between stellar mass and metallicity.
Recently, Tremonti et al. (2004) confirmed the existence
of the stellar mass-metallicity relation (MMR) on statistical basis
for a sample of galaxies in the Sloan Digital Sky Survey 
(SDSS). 
These authors detected a change in the slope of this relation at a
 stellar mass similar to the characteristic mass defined by
Kauffmann et al (2003).
Tremonti et al. (2004) suggested that the steeper slope found in this relation   for
smaller stellar masses was the result of the action of galactic winds.
 Recently, Gallazzi et al. (2005) found that stellar mass is not the fundamental parameter 
determining
age or metallicity based on the large dispersions found in their data.
The physical origin of both  the luminosity-metallicity and the mass-metallicity relations,
 their interconnection and  evolution  are  fundamental problems for
models of galaxy formation.

High precision, large-scale cosmological observations provide strong support for $\Lambda$CDM models.
In this scenario, galaxies form by  the hierarchical aggregation  of
substructure. Hence, mergers and interactions as well as  continuous
gas inflows play a crucial role in the history of evolution of
galaxies, affecting the mass distributions and star formation
rates, and as a consequence, 
the chemical properties of the systems.
In order to study galaxy formation and  to provide a consistent
concatenation and 
interpretation of   observational results, 
sophisticated models which can follow the hierarchical building up of
the structure together with the evolution of their chemical properties are needed.
Cosmological hydrodynamical models which include chemical production 
 are a powerful tool to  tackle this problem as it has been previously
shown 
by  Mosconi et al. (2001) and Lia et al. (2001), among others.

%In this Letter, we use the chemodynamical Gadget-2 of Scannapieco et al. (2005)
%to study the formation of galaxies and the origin of (stellar) mass-metallicity and the velocity-metallicity relations
%in a  hierarchical clustering scenario. Our numerical simulations include star formation, metal-dependent
%radiative cooling and chemical enrichment by Supernova II and Ia. Energy feedback is not treated 
%in this work. 
%However, discussions on its possible effects  are included.

The main goal of this work is to discuss the direct relation between the dynamical evolution and chemical properties
of galactic systems
in a hierarchical clustering scenario,
in order to provide clues on the origin of the fundamental metallicity relations of galaxies.
 Note that this paper is focused on the study of systems with stellar masses larger than
$10^9 {\rm M_{\odot}}h^{-1}$.
Hence,results may not be  relevant for  dwarf galaxies, where SN are expected to play a more important role
Although we do not include a strong Supernova (SN) energy
feedback in this paper, we discuss  its  potential action on
the  results.

This Letter is organized as follows. Section 2 describes the numerical simulations, the methodology applied to
estimate the chemical and astrophysical properties of galactic systems and the findings.
Section 4 summarizes the main conclusions.

\section{Analysis and Results}

We have run cosmological hydrodynamical simulations which describe the evolution of dark matter and baryons
in a $\Lambda$CDM scenario ($\Omega=0.3, \Lambda =0.7, \Omega_b =0.04$
and $H_0 = 100 \ h \ {\rm km\  s^{-1}Mpc^{-1}}$, $h=0.7$)
including metal-dependent radiative cooling, star formation and chemical enrichment by Supernovae II and Ia. 
The chemical 
algorithms  have been developed within the code
 {\small GADGET-2} in its fully conservative version \cite{scan05}.
 The simulation  corresponds to a  $10 \ {\rm Mpc } \ h^{-1}$ side cubic volume 
resolved  with initially $2\times 160^3$ 
 particles, implying a mass 
resolution of  $1.7 \times 10^7 \msun$ h$^{-1}$
 and  $2.0 \times 10^6 \msun$ for the dark matter and gas,
 respectively.
We have assumed an instantaneous thermalization of the supernova (SN) energy.

Virialized structures are identified at different redshifts 
by using a combination of a friends-of-friends technique and
a contrast density criterium ($\delta \rho/ \rho \approx 178 \Omega^{-0.6}$).
The dynamical and mean chemical properties of these systems,
as well as their star formation histories, are estimated taking into
account the mass within
the optical radius, defined as the one that encloses 
$83 \%$ of the baryonic mass.
Since these simulations follow the chemical enrichment by  individual
elements,  we can directly estimate
abundance indicators such as  ${\rm 12 + log (O/H)}$. 
We have chosen to work with mass-weighted abundances since they
reflect more reliably the chemical enrichment of the system as a
whole. 
Caution should be taken when comparing these abundances with  observations that provide information only 
on certain regions of galaxies  such as HII regions.

We have constructed the stellar mass-metallicity relation and the circular velocity-metallicity relation
for galactic systems at different redshifts.
The circular velocities ($V_{\rm opt}$) are measured at the optical radius.
For the gas component, we find that  the scatter in the relations is much larger  and their evolution
 is not that clearly defined, although the gas follows the
same general trends defined by stars (see De Rossi et al. 2005, in preparation).
For that reason, in this Letter we estimate the metallicities from the simulated stellar
populations.

\begin{figure}
\begin{center}
\vspace*{-0.05cm}\resizebox{7.5cm}{!}{\includegraphics{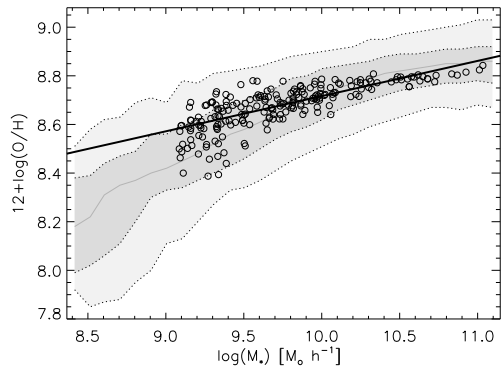}}\hspace*{-0.2cm}%
%\resizebox{4.5cm}{!}{\includegraphics{fig2b.eps}}\vspace*{-0.24cm}\\%
%\resizebox{4.5cm}{!}{\includegraphics{fig2c.eps}}\hspace*{-0.2cm}%
%\resizebox{4.5cm}{!}{\includegraphics{fig2d.eps}}\vspace*{-0.2cm}\\%
\end{center}
\caption{Mass-metallicity relation for the simulated galactic systems
at $z=0$ in a $\Lambda$-CDM scenario (open circles).
The shaded areas correspond to one and two $\sigma$ regions 
from the observed relation of Tremonti et al. (2005).
We have also included a linear regression fit to the simulated points for comparison (solid line).
}
\label{zmoh}
\end{figure}

In Fig.~\ref{zmoh} we show the MMR estimated for the galactic systems 
at $z=0$, renormalized by 0.25 dex to match the zero point of the observed
relation given by  Tremonti et al. (2004; shaded areas). We
note that this displacement is not needed for
the gas component which is always more enriched than the overall stellar population.
However, we would also like to point out   that the  metallicities corresponding to the SDSS have
been obtained from 
the central regions of galaxies which are expected to be more metal-rich. 
 As it can be seen from
this figure, a linear regression fit  is not a good representation of  the
simulated relation. However, the simulated MMR is in agreement with the trend
estimated from the SDSS, although  there is  some excess of metals at the lower mass end.

\begin{figure}
\begin{center}
\vspace*{-0.5cm}\resizebox{6.5cm}{!}{\includegraphics{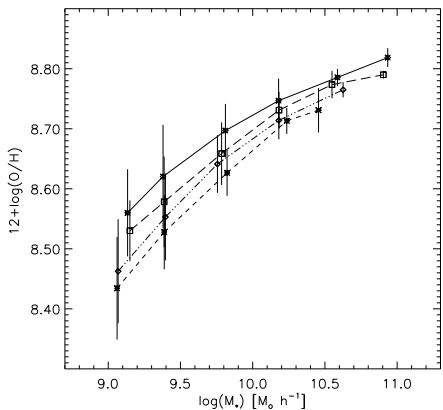}}\hspace*{-0.2cm}\\
%\resizebox{6.5cm}{!}{\includegraphics{fig2b.eps}}\hspace*{-0.2cm}%
%\resizebox{6.2cm}{!}{\includegraphics{fig3new.eps}}\hspace*{-0.2cm}%
\end{center}
\caption{ Mean mass-metallicity relation 
 for  galactic systems identified at 
 $z=3$ (short-dashed lines), $z=2$ (dotted-dashed lines), $z=1$ (long-dashed lines) and $z=0$ 
(solid lines) in a $\Lambda$-CDM scenario . Error bars correspond to the standard rms dispersions.
}
\label{zmstar}
\end{figure}

\begin{figure}
\begin{center}
\vspace*{-0.5cm}\resizebox{7.0cm}{!}{\includegraphics{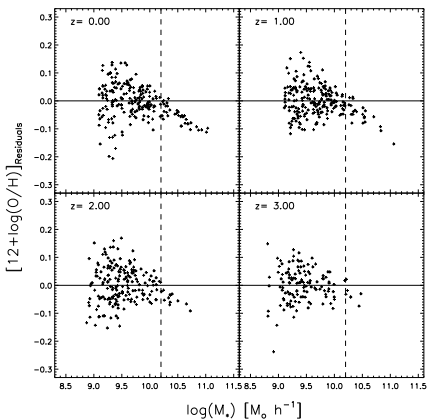}}\hspace*{-0.2cm}
%\vspace*{-0.5cm}\resizebox{6.5cm}{!}{\includegraphics{fig2b.eps}}\hspace*{-0.2cm}
\end{center}
\caption{Residuals of 12+log (O/H) with respect to a linear regression fit to the relation
between metallicity and stellar mass at redshifts of  Fig.~\ref{zmstar}. The dashed lines show
the stellar mass at which the residuals are systematically negative because
of the change in the slope of the stellar mass-metallicity relation. 
}
\label{fig3new}
\end{figure}

In Fig.~\ref{zmstar} we show the stellar mass-metallicity relation
for galactic systems at different redshifts. We  can appreciate that the MMRs
behave in a similar fashion from $z=3$ to the present time. 
As it can be seen from this figure,
there is a       change in the curvature at 
$M_c \approx  10^{10.2} \msun h^{-1}  $ 
which corresponds to an abundance of 
$12 +  {\rm log (O/H)_c} \approx   8.7$. 
This characteristic mass has been estimated by determining 
the stellar mass where the linear fit is no longer
a good representation of the simulated data. This has been achieved  by  the analysis  of the residuals of these
relations. We estimated the stellar mass at which the residuals show a behaviour
which departures from that expected for a good  linear fit. As it can be seen in Fig. ~\ref{fig3new},
for $M_* \ge M_c$, the residuals start
to be systematically negative indicating a saturation of the metallicity as a function of
stellar mass. 
For all analysed redshifts, the relations determined by the systems with
stellar masses greater than $M_{\rm c}$ tend to have a shallower slope,
although the level of enrichment increases with redshift.
Conversely, systems with $M_*<M_{\rm c}$ show a steeper correlation between  
metallicity  and stellar mass.
The error bars in  Fig.~\ref{zmstar}
denote the statistical dispersion of  the abundances in
each bin.
These dispersions show a larger scatter in  oxygen abundances
for $M_* < M_{\rm c}$ which suggest that  the  histories of evolution of these systems are more 
different among each other than those of the systems at the massive end (a similar trend
is found by binning in mass intervals with equal number of members).

The redshift evolution of the MMR relation indicates an increase with time 
of the chemical content of the systems,
with the major changes  driven by the smaller ones which, on
average,  are enriched by up to $\approx 0.10$ dex
from $z \approx 3$ to $z=0$. This behaviour is responsible of a general flattening of
the slope in this mass range.
  Systems with  $M_* > M_{\rm c}$ show less
evolution in their chemical content with a $\approx$0.05 dex variation from $z=3$ to $z=0$. 
The characteristic mass remains almost unchanged with redshift (Fig. ~\ref{fig3new}) 
while its corresponding oxygen  abundance 
increases only by 0.05 dex in the same redshift range.  

The different histories of evolution of the systems 
may be determining the dependence of the metallicity evolution and
 dispersions on  stellar mass. 
In order to investigate this point,  we estimated
the time ($\tau_{50}$) when $50\%$ of the total  stellar mass at $z=0$
was already formed for each galactic system.
This time $\tau_{50}$ is defined by 
the particular evolutionary history of each galactic system. As expected, we found that objects
with $M_* > M_{\rm c}$ have older stellar populations  with a
 variation of up to  $\approx 15 \%$ in $\tau_{50}$. Conversely,
the stellar populations of the smaller mass systems present a wider
 range of ages,  leading to a change of more than $ 30\%$
in $\tau_{50}$.
Hence, while most of the stellar content of massive systems is formed at 
higher redshift as expected in this kind of cosmological scenario,
smaller systems constitute a diverse population.

In order to understand the physical meaning of the evolution of the
MMR and the 
characteristic mass $M_{\rm c}$,
 we have analysed the merger trees of the simulated systems at $z=0$ 
and the
relation between the dynamical and chemical properties of their
progenitors at $z>0$ (De Rossi et al. 2005, in preparation).
From the  analysis of the merger histories we find that systems with
$M_* <M_{\rm c}$ 
transform their gas content into stars in a more passive
fashion or via gas-rich mergers,  setting
a steeper correlation between stellar mass and  metallicity.  
Galaxies with masses larger than 
  $M_{\rm c}$ are formed by  merger events which involve stellar 
dominated systems. 
In these cases,  the mergers  produce a system with a final stellar mass
equal to the sum of the old stars  in the merging objects plus some new born ones,
while its overall mean abundances remain at the same level of enrichment.
This situation
occurs because, in this case, the merging systems have already transformed most of their gas into
stars so that there is no  fuel for an important starburst during the merger.
It could be also possible that a large system merges with a smaller less-enriched one 
which can feed  new star formation activity but with lower level of enrichment.
Both scenarios have the same flattening effect on the slope of
the mass-metallicity relation in 
systems within this range of masses.

\begin{figure}
\begin{center}
%\vspace*{-0.5cm}\resizebox{7.5cm}{!}{\includegraphics{fig3new.eps}}\hspace*{-0.2cm}
\vspace*{-0.5cm}\resizebox{6.5cm}{!}{\includegraphics{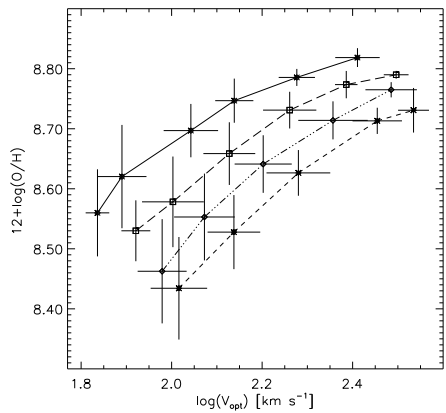}}\hspace*{-0.2cm}
\end{center}
\caption{ Mean  circular
velocity-metallicity relation
 for  galactic systems at the same redshifts shown in Fig.~\ref{zmstar}.
  Error bars correspond to the standard rms dispersions.
}
\label{fig4}
\end{figure}

We  also calculated the  circular velocity -metallicity
relation   as shown in Fig. ~\ref{fig4}. 
We used the same mass intervals  as in  Fig.~\ref{zmstar} in order 
to estimate the mean velocity of the corresponding galactic systems and
their standard deviation in each mass interval.
We can appreciate from this figure that, at a given redshift, the 
faster a system rotates, the higher its metallicity.
At a given velocity, the chemical evolution with redshift   is
larger than that obtained at a given  stellar mass. 
This is because as   one moves to lower redshift, more massive systems start to contribute to the lower circular velocity intervals.
 At high
redshift, galactic systems are more concentrated with less mass required to get to higher velocities.
This reflects the fact that
as the Universe expands and its mean density decreases, galactic systems do not need to reach so high densities to 
form bounded structures. 
Although the circular velocity determines a clear correlation, it shows larger evolution with time
than the MMR.
 If the circular velocity of a system is known, the redshift of the systems is also 
needed to  establish its metallicity, otherwise
the dispersion could be as large as $\approx 0.35$ dex.

In the simulations analysed in this Letter, 
we have not included a treatment of energy feedback, 
which is expected to produce powerful outflows, affecting
the star formation process and the metal production and mixing. 
Hence, the treatment of energy feedback may introduce 
changes in  the metallicity properties. 
Taking into account previous results, SN feedback is expected to
affect more strongly  systems with circular velocities  smaller than $100 \kms$ (Larson 1974).
In this case, only the smallest
systems  in Fig.~\ref{fig4} would be significantly affected. 
Note that the velocity corresponding to the characteristic  mass
varies from around $300 \kms$ at $z=3$ to $ 140 \kms$ at $z=0$ and consequently, it is unlikey that $M_{\rm c}$
will be significantly modified either. 
At $z=0$, the strongest effects of energy feedback are expected to
take place in the small mass range 
where we detect an excess of metals (see Fig~\ref{zmoh}). The action of strong SN   feedback 
would produce  the ejection of  part of the enriched material out of the systems and the  decrease of
the  general  level of enrichment.

The need for  SN outflows is also suggested by the analysis of the effective yields
(defined as $y_{Z}= Z /{\rm ln} (\mu)^{-1}$ where $Z$ is the gas-phase metallicity and $\mu$ the gas fraction
of the systems). 
 The   simulated $y_{Z}$ values are lower than the solar yield
expected in a closed box model, since our systems formed in a hierarchical scenario where mergers, interactions and infall
affect the mass distribution and regulate star formation.
Contrary to observations (e.g. Garnett 2002), we found slightly larger $y_{\rm Z}$ for systems with $10^9 {\rm M_{\odot}} h^{-1} < M_* < M_{\rm c}$
 compared to those of the massive ones ($ M_* > M_{\rm c}$),
 which supports the claim for stronger SN outflows for the former. 
In the case of massive systems, we get
a mean  of $y_{Z}$ which remains approximately constant with optical velocity.
Although this latest result is in agreement with Garnett (2002), it might be  indicating 
 the need for some ejection of  material also  in massive systems in the light of
the new observational  findings of  Tremonti et al. (2004).

\section{Discussion and Conclusions}

We have  shown that galactic systems in the concordance $\Lambda$CDM  model  reproduce
a  stellar mass-metallicity relation in general agreement with that  measured for galaxies 
in  the SDSS at $z\approx 0$,
as a result of the particular evolution of  galactic systems in this kind of scenarios. 
 We acknowledge an excess of metals at the low mass end which could
be solved by strong SN energy feedback.

We found a stellar mass-metallicity relation well-defined from $z=3$ which evolves weakly  with time. 
The largest change in the metal content is driven by the smaller mass systems.
We have also determined  a characteristic stellar  mass (corresponding to  solar  abundance)
so that for more massive systems,   the mass-metallicity relation flattens and
there is  less evolution in their chemical enrichment level. 
This characteristic mass and its corresponding metallicity are robust
against  numerical resolution since approximately the same values are
detected in experiments run with
different number of particles, from $50^3$ to $160^3$.

The characteristic mass found from our simulations agrees with the one 
estimated from the SDSS by Kauffmann et al. (2003).
This mass appears naturally as a result of the hierarchical building up of
the structure  and segregates two distinctive types of galactic systems.
Our findings show that systems with $M_* >M_{\rm c}$ transform most of
their gas into stars at higher redshifts
and experience important merger events. At lower redshifts, these mergers tend to involve  
 systems dominated by stars so that while the final stellar mass of the outcoming system is larger,
 its mean stellar  metallicity remains
 basically the same. This is because a small percentage of new stars are formed during the merger and,
 consequently, the final metallicity is determined by the old stars belonging to  the incoming systems.
%The effective yields of these systems are subsolar and consistent with those
%derived from the SDSS by Tremonti et al. (2004).
Smaller galactic objects form their stars in a more passive way or
during  gas-rich mergers. 
Hence, in this case, there is a more steeper correlation between metallicity and stellar mass content. 
%We also
%find that the non-linear building up of the structure produce a depletion  of gas into stars given rise to a relation
%between stellar mass and gas fraction in agreement with observations (Boselli et al. 2001). 

According to our results,  in the absence of strong SN energy feedback,
the stellar mass-metallicity relation is the parameter that better determines
the metallicity of a galaxy with a weak dependence on redshift.
Conversely, at a given circular velocity, the combination of chemical enrichment
and cosmology produces a larger evolution with lookback time.
% Conversely, the cosmological 
%evolution of circular velocity-metallicity
%relation is larger with  cosmology playing an important role.
We found that the dispersions in both relations
 are produced by the combination of stars with different ages and metallicities 
which results from  the particular dynamical history of evolution of a system
 in a hierarchical clustering scenario.
%which establish the characteristics  of the stellar populations.

Acknowledgements. 
We are grateful to the anonymous referee, CONICET, 
 Fundaci\'on Antorchas and LENAC.
The simulations were
performed on the Ingeld PC Cluster   and
on HOPE cluster  at
IAFE (Argentina).

\bibliographystyle{apj}
%\bibliography{letter}

~

\end{document}